\documentclass{article}

\usepackage{PRIMEarxiv}

\usepackage[utf8]{inputenc} 
\usepackage[T1]{fontenc}    
\usepackage{hyperref}       
\usepackage{url}            
\usepackage{booktabs}       
\usepackage{amsfonts}       
\usepackage{nicefrac}       
\usepackage{microtype}      
\usepackage{lipsum}
\usepackage{fancyhdr}       
\usepackage{graphicx}       
\graphicspath{{media/}}     

\title{Anonymous Expression in an Online Community for Women in China
}

\author{
  Zhixuan Zhou \\
  University of Illinois at Urbana-Champaign \\
  \texttt{zz78@illinois.edu} \\
   \And
  Zixin Wang \\
  Zhejiang University \\
  \texttt{zixin\_wang@zju.edu.cn} \\
   \And
  Franziska Zimmer\thanks{Corresponding author.} \\
  Heinrich Heine University Düsseldorf \\
  \texttt{franziska.zimmer@hhu.de} \\
}

\begin{document}
\maketitle

\begin{abstract}
Gender issues faced by women can range from workplace harassment to domestic violence. While publicly disclosing these issues on social media can be hard, some may incline to express themselves anonymously. We approached such an anonymous female community on Chinese social media where discussion on gender issues takes place with a qualitative content analysis. By observing anonymous experiences contributed by female users and made publicly available by an influencer, we identified 20 issues commonly discussed, with cheating-partner, controlling parents and age anxiety taking the lead. The results are placed into context with Chinese culture and expectations about gender.
By describing the results in context with the social challenges faced by women in China, and understanding how these issues are anonymously and openly discussed by them, we aim to motivate more policies and platform designs to accommodate the needs of the affected population.  
\end{abstract}


\section{Introduction}
Gender issues discussed in online spaces include workplace harassment, intimate partner violence (IPV) and misogyny, among many others \cite{weibo:1, introduction:1}. Talking about these issues openly can help the people affected by such circumstances to seek help or at least feel understood. In Western cultures, there seems to be an increasing trend to post these stories non-anonymously, as prevalent by the \#MeToo movement on social media \cite{introduction:1,introduction:2}. 

Nevertheless, the situation can be especially dire for women in certain countries where familial issues like IPV are not openly talked about, e.g., in China, where face-saving and shame can hinder open discussion \cite{weibo:1}. This does not mean that these issues do not exist in Mainland China. A study focusing on this region estimates that the victimization of IPV in the general population can be estimated to be around 17.4\% to 24.5\% for psychological violence, 2.5\% to 5.5\% for physical violence, and 0.3\% to 1.7\% for sexual violence. Furthermore, it was noted that studies on IPV are relatively rudimentary and have had a low priority for research about Chinese society \cite{IPV:1}. There seems to be a lack of empirical studies investigating the risk factors associated with IPV as well \cite{IPV:2}. 

However, recently, it seems increasingly common for Chinese women to anonymously reveal their (mostly negative) experiences such as unfair treatments by partners or mental issues to Sina Weibo influencers, and the influencers would post their experiences which can be seen by the followers. Some typical examples include, e.g., being forced to get married early by one's parents, one's cheating behavior as a retaliation to her partner's cheating behavior, misogynistic speech by co-workers. Such topics can strike a response in other women, and a discussion would take place in the comments, containing both empathy or practical advice, but also, personal attack or moral judgment. 

Toward understanding how gender issues are anonymously discussed by women on Chinese social media, an exploratory content analysis by observing 100 anonymous experiences shared with a female influencer was conducted. Twenty gender issues were identified in the anonymously expressed experiences, including, e.g., harassment, lack of emotional support from partners, cheating, domestic violence—most of which are associated with intimate relationships. It was also observed that some women seem to show tolerance towards their partners, as well as to attribute the issues in relationships to themselves. To the author's knowledge, this is the first empirical understanding of how gender issues are anonymously discussed by women on Chinese social media. 

Following, related work on anonymous online expression and female online communities will be outlined, and the research questions proposed. Then the methodology is described and results reported. Summarizing main takeaways as well as suggesting future directions conclude this study.

\section{Related work}
There are two lines of research closely related to this study: anonymous online expression and online communities for women. Anonymous expression on the Internet describes the ability to interact online without having to use identifying markers, e.g., birth names or age \cite{anonymous:7}. Sometimes, women can feel more confident in sharing their experiences in online spaces designed for or frequented by them, anonymously or not.

\subsection{Anonymous online expression}
Anonymity is a preferred feature of online communication at times, which could encourage expressiveness and interaction among users, and allow more honesty, openness, and diversity of opinion \cite{anonymous:7}. People share various types of content in anonymous communication applications, ranging from deep confessions and secrets to lighthearted jokes and momentary feelings. Important motivations for participation and posting are to get social validation from others, even though they are anonymous strangers \cite{anonymous:7}, or just out of boredom and for fun \cite{anonymous:9}. Although researchers often regarded user identity and data permanence as central tools in the design of online communities, a study of 4chan, an anonymous English-language imageboard website, found that over 90\% of posts were made by fully anonymous users \cite{anonymous:1}. 

Anonymous online communities make online spaces valuable for users, especially regarding sensitive topics. When the state cannot provide a particular service, people may turn to the Internet to find alternative information and social support. For example, a case study of a Russian Alcoholics Anonymous (AA) online group uncovered how people sought help for their drinking problems \cite{anonymous:2}. Members were less likely to enact ``unidentifiability'' if they were more connected to the particular community and had more time in recovery \cite{anonymous:6}. 

Favoring anonymity may be positively correlated with narcissism and low self-esteem and research suggests that users with stronger anonymity preference tended to be younger, highly trusting, having strong ties to online communities while having few offline friends \cite{anonymous:4}. There is also research suggesting that anonymous sites such as Reddit are more often used by male than female users. In the case of Reddit, citizens of the United States represent the largest user group \cite{reddit:1, reddit:2}. Further, male and female users enjoy using Reddit to the same extend and see it as easy to use, enjoyable and informative \cite{reddit:3}.

With anonymity on social media there also come potential problems, such as prevailing sexual harassment, bullying, or even illegal acts \cite{anonymous:8}. An analysis of YikYak, an anonymous, location-based social media smartphone application, found frequent occurrences of profanity, vulgarity and sexual references. However, the potential for abusive postings may be mediated by online community policies \cite{anonymous:3}. 

Online communication, whether anonymous or not, may breed socio-demographic homophily. By analyzing behavioral data of 7,287 users of a Korean anonymous online dating advice platform, Kang and Chung found that one was more likely to respond to problems submitted by advice seekers of a comparable age, and advice seekers were more likely to approve of a response if the advice seeker and advice provider had similar educational backgrounds \cite{anonymous:5}.

Focusing on research conducted in China and Chinese online spaces, the most popular microblogging service is Sina Weibo, a service unique to the country, which is more often used by female than male users \cite{weibo:3}. Weibo is described as a platform for maintaining friendships, recording personal lives, expressing oneself, and gathering information in terms of social issues or personal interests \cite{weibo:4}. Themes that are discussed on Weibo include depression, philosophical thoughts on life, medical information, and help seeking \cite{weibo:5}. Users on Weibo tend to mainly share positive self-disclosure and fewer negative information, especially if the person is less anonymous \cite{weibo:2}. Nevertheless, the combination of anonymity and present opinion leaders on Weibo or other social media helps people to share their experiences (e.g., about IPV). The anonymity feature of social media has also mitigated the negative effect of some traditional mentalities, such as family shame and face-saving, and encouraged more expression and discussion \cite{weibo:1}. 

\subsection{Online communities for women}
There are some social networking sites exclusively, mainly, or initially designed for women. For example, CafeMom\footnote{\url{http://www.cafemom.com/}} is a social networking site for moms to discuss parenting issues. POPSUGAR\footnote{\url{https://www.popsugar.com/}} is another one for young women to discuss latest gossip, play games, and share tips with each other. The success of Xiaohongshu\footnote{\url{https://www.xiaohongshu.com/}}, a Chinese social media and e-commerce platform, has been known to stem from its emphasis on consumption of media associated with women. Its culture concept, media convergence and marketing are highly connected with feminism \cite{community:4}. Chinese moms actively vlog on Xiaohongshu, creating a community for self-expression and mutual support \cite{mom}. 

Early research of the formation of communities for women was mostly about entertainment and daily life. For example, an ethnographic study of two Internet communities, one for fans of the American television series The X-Files, and one for the Canadian series Due South, explored fandoms, and examined negotiations of gender, class, sexuality and nationality in making meaning out of a television show \cite{community:1}. MissyUSA\footnote{\url{www.MissyUSA.com}} is an online community of Korean immigrant women in the United States. These women's shared identity (i.e. being Korean, married, female, and living in the United States) was found important in the formation and development of this online community. In this candid talking space, they vent their innermost feelings about their lives in the United States. Information sharing and help offering were also found in this community \cite{community:2}.

This study further focuses on a community for women on Sina Weibo initiated and maintained by a woman influencer who identifies as a feminist \cite{community:5}. Similar to the study on anonymous posts on Weibo about IPV \cite{weibo:1}, it needs to be investigated what other topics are anonymously discussed in Chinese online spaces, especially by women. Therefore, the research questions are:
\begin{itemize}
    \item \textbf{RQ1: }What are common gender issues revealed in anonymous expression of female Weibo users? 
    \item \textbf{RQ2: }How can the observed issues be contextualized in Chinese cultures? 
\end{itemize}

\section{Methods}
Some Weibo users disclose their personal experiences to well-established influencers with thousands or even millions of followers, expecting their experiences to be posted in de-identified images, often for expression and advice seeking purposes \cite{Z:young}. Based on this observation, we conducted an exploratory study of the anonymous disclosure of gender issues on Weibo under a specific feminist influencer, called Qiaomai (\begin{CJK}{UTF8}{gbsn}荞麦\end{CJK}), for several reasons. First, female users seem to prefer Weibo in contrast to male users, resulting in potentially higher number of posts \cite{weibo:3}. Second, she had over 1.5M followers as of April 2022, and a majority of them were female users. With mainly self-disclosed experiences and comments reflecting women's views, the issues at hand could be better understood. Third, as a feminist herself, Qiaomai made an effort to create a supportive community for women. 

In the end, 774 posts from January to March 2022, as well as corresponding comments, were collected. Due to the time limit, only the first 100 of the anonymous self-disclosures were used for analysis in the current study. Since they were in the form of images, the text was extracted using an automatic OCR tool\footnote{\url{https://saas.xfyun.cn/ocr?ch=sa02}}, and mistakes were corrected manually.


\subsection{Content analysis}
A content analysis was applied because it can be challenging to obtain detailed events or expressions of sensitive subjects when using interviews or questionnaires. This is especially relevant to Chinese culture where face-saving regarding sensitive topics such as IPV can hinder open disclosure \cite{weibo:1}. To our knowledge, there was only one prior study about Chinese women sharing their experiences (anonymously) on social media \cite{weibo:1}. Thus, we applied a conventional approach to content analysis \cite{method:1}, and let the themes emerge during the analysis.

Two authors independently coded the posts, and ensured reliability by calculating Krippendorff's alpha \cite{method:2}. The research team regularly met and discussed to refine the coding. After a first round of content analysis and categorizing posts, the codebook was updated to guarantee a good inter-coder reliability. Each post was observed for corresponding content categories: if a category was applicable, it was marked accordingly, resulting in a nominal data set. 

The categories that emerged after using the conventional content analysis approach by observing 100 posts were: ambiguous relationship; positive experience; reflection on relationship, life, etc.; health issue; harassment; workplace issue; gender stereotype/reproductive freedom; parenting; love vs. money; betrothal gift; debts/gambling; domestic violence; lack of emotional support/romance; sex mismatch; concealing of illness/family issue; work-family balance; cheating-self; cheating-partner; age anxiety/marry early; controlling parents/dependence. In addition to the gender issues above, we also observed two mentalities in the self-disclosure, namely, tolerance and self attribution. 

In the subsequent results section, occurrences of the categories will be reported. Quotes from the self-disclosures will be used to contextualize each gender issue. The self-disclosures are all anonymous, thus there will not be ethical concerns regarding anonymity violation. 

\begin{table*}[h]
\centering
\caption{Descriptive statistics of content categories. Sometimes multiple assignments.}
\label{statistic}
\begin{tabular}{l|rl}
\hline \bf Content Category & \bf Occurrence & \bf Krippendorff's Alpha \\ \hline
ambiguous relationship & 8 & 1 \\
positive experience	& 14 & 1 \\
reflection on relationship, life, etc. & 8 & 1 \\
health issue & 3 & 1 \\
harassment & 1 & 1 \\
workplace issue & 5 & 1 \\
gender stereotype/reproductive freedom & 10 & 0.879 \\
parenting & 3 & 1 \\
love vs. money & 3 & 0.653 \\
betrothal gift & 1 & 1 \\
debts/gambling & 3 & 1 \\
domestic violence & 8 & 0.936 \\
lack of emotional support/romance & 12 & 0.906 \\
sex mismatch & 5 & 1 \\
concealing of illness/family issue & 1 & 0.663 \\
work-family balance & 8 & 0.826 \\
cheating-self & 11 & 0.889 \\
cheating-partner & 19 & 1 \\
age anxiety/marry early & 14 & 0.96 \\
controlling parents/dependence & 16 & 0.922 \\ \hline
tolerance & 10 & 0.756 \\
self attribution & 5 & 0.754 \\
\hline
\end{tabular}
\end{table*}

\section{Results}
Based on the content analysis, anonymously expressed issues experienced by women and their occurrence frequency could be identified. By presenting quotes from the posts, nuanced context of each issue can be vividly displayed.

\subsection{Descriptive statistics}
Among the 100 posts that were annotated, the majority of them were self-reported experiences (N=83). The rest were either narration of other people's experience, or a combination of self-narration and narration of other people. Notably, 99 of the 100 posts analyzed were contributed by self-reported female Weibo users, confirming the assumption that the site for analysis of our choice is indeed a community for women. 

A large range of issues shared by women could be identified during the annotation process, with positive experience, gender stereotype/reproductive freedom, lack of emotional support/romance, cheating-self, cheating-partner, age anxiety/marry early, controlling parents/dependence as the most prominent ones (N\textgreater10). The occurrence of each issue can be found in Table~\ref{statistic}. It is worth noticing that similar themes may appear together, as the sharing of one experience may motivate the sharing of more similar experiences. 

The two coders reached a high agreement on most issues with Krippendorff's Alpha larger than 0.8 (\cite[p.~241]{method:2}. The rare exceptions are with love vs. money and concealing of illness/family issue. The overall Krippendorff's Alpha averaged 0.92 on all variables, demonstrating a high degree of agreement between the two coders. 

A non-negligible percent, i.e., 10\%, of the content contributors showed, in the eyes of the authors, a lot of tolerance for their partners. They may nevertheless describe their partners as having many strengths even if they were treated badly. Further, 5\% of the content contributors attributed the issues in the relationship to themselves, even though it is not obvious who caused the issue, at least in the author's eyes. 

\subsection{Contextualizing gender issues}
In posts regarding \textbf{ambiguous relationship}, one may confess her mental struggle or reflection on early stage relationships such as dating or blind dates. For example, one woman hesitated if she should establish a more ``formal'' relationship with a newly met man, ``\textit{I may look very independent, but when I meet someone I'm fond of, I would want to rely on him, though I'll just hide this feeling. This may cause misunderstanding. I'm wondering if I should have more courage, and express my true feelings from inner inside. This is really hard for me. Hope I can make some progress in the new year.}'' (Post 3). 
As could be seen, such posts may overlap with the ``reflection on relationship'' category. Ambiguous relationship can also relate to cheating behavior, given the ambiguous boundary between cheating and friendship, as in the case of Post 42: ``\textit{I happened to see his chat history with a colleague of his. Though there's not much loving content, but I can feel their relationship which goes beyond that of friends in their obviously overly frequent communication.}''  

The sharing of \textbf{positive experience} often follows the narration of a toxic relationship or marriage, e.g., ``\textit{After her divorce, my mom often travels to neighboring cities. She also starts to learn dancing, square dancing though. She becomes more relaxed.}" (Post 2). Post 66 describes a similar case, ``\textit{The most courageous thing I've done in my life is to get a divorce. Now I work hard and get along well with my family and friends. I believe my son will grow happily with his happy, hard-working mother... It's not scary to make choices, but we must have the courage to face reality after making the choices.}''
Other themes in this category include academic achievement (Post 14) and satisfactory relationship/partner (Post 31, 51, 69, 75, 78, 88, 89, 97).

\textbf{Reflection on relationship, life, etc.} is commonly found among the posts, including whether or not to go into a marriage (Post 11, 24, 26), the difficulty of finding satisfactory partners (Post 81), and how to live an independent life (Post 93). This category often does not contain any specific life events, while providing reflection of the women.

\textbf{Health issue} is relatively rarely shared. Two posts are about the discrimination faced by Hepatitis B virus carriers, in both workplace and life: ``\textit{My mom is a really nice person. She faces much discrimination due to her disease. People don't want to be around her. She's isolated by people, but she still constantly strives to become stronger.}'' (Post 85). According to Post 68, the content contributor intentionally isolates herself from others, out of fear for letting others know about her disease and infecting her friends.

\textbf{Harassment} is mentioned only once, where a woman was sexually harassed when she was a child (Post 9).

Severe \textbf{workplace issues} were also described. In Post 1, the woman explains how she was verbally abused by a sexist male client: ``\textit{The male client proposed a toast. I said to my colleague next to me, `I'm drunk. I don't want to drink anymore.' The client may have heard my words and said, `I'd like to marry every beauty here, except her. Do you know why? Because she's not obedient and doesn't give me face (do me a favor). When other people propose a toast, she drinks. When it comes to me, she doesn't.' Then the client kept repeating that I did not obey.}'' Post 56 tells how a woman strives to leave a toxic workplace. Workplace issue is sometimes related to a toxic relationship, as in the case of Post 66: ``\textit{I became weak and self-contemptuous in the relationships. I couldn't see my strengths. When I brought such emotions to the workplace, I was bullied and criticized  by leaders.}''

\textbf{Gender stereotypes} are an umbrella of many issues related to gender roles, with the lack of reproductive freedom as the most prevalent one. The author of Post 21 describes how her father-in-law forced her to have a baby, regardless of her health conditions: ``\textit{His father has the final say in his family. It's a very traditional, patriarchal family. I don't know how many times I've gone to the hospital, in order to have a baby. I've heard much bad words from his parents. Now his father is like, `either you manage to have a test tube baby, or you divorce.'}'' Women are the ones being responsible for housework and taking care of the children, without any help from their partners, as complained about in Post 46 and 70. Gender stereotypes can also relate to traditional mindsets. For instance, some parents do not allow their married daughter to come home on Lunar New year's Eve and first day of the lunar year, because they believe ``\textit{this would bring bad fortune to her family.}'' (Post 25). 

\textbf{Parenting} issues are raised when differing parenting principles occur between parents. Two posts talk about how bad men seem to be at parenting. In Post 62, the author says, ``\textit{Sometimes I really think men, especially after they become fathers, are good-for-nothing. They're lazy and ugly, but keen to show the so-called father majesty. In fact, they're just incompetent and stupid.}'' The author of Post 71 mentioned that her partner was impatient to their child, and would throw cellphones and feeding bottles.

Sometimes, \textbf{love vs. money} is the choice some women have to make when choosing between material standards of living and a romantic relationship. In China, traditionally, men are responsible for down payment of the house when they get married. The content contributor of Post 4 thinks about breaking up with her boyfriend who is not willing to pay the down payment: ``\textit{Now whenever we talk about getting married and buying the house, he says he's not able to do it and wants to give up. I just want to ask, am I asking for too much? Should I keep making compromises? Is it the problem with his family?}'' Post 67 is also written by a woman who thinks her boyfriend does not spend enough money on her: ``\textit{I think he's not spending enough money on me. I'm young, and pretty. I have a good educational background. Also, he's more than ten years older than me. So, he should devote more to provide me with a good material life.}'' The life of marrying a rich partner, however, is not all fantasy. In Post 60, the content contributor was asked to sign a prenuptial agreement. She felt sad after she realized that her husband and parent-in-laws guarded her in terms of money: ``\textit{Sometimes I feel that I have nothing to complain about. My family of origin has no money. With my own efforts, I can't live at this level of life. What else can I expect? But sometimes I feel very sad. While they are good to me, they're also guarding me in terms of money, and they will not let me take a little advantage.}''

One post (Post 4) touches on the ancient tradition of \textbf{betrothal gift} which is a traditional way of thanking and showing respect to the bride's parents for raising the bride, and a desire to forge good relations with the future parent-in-laws. In this post, the betrothal gift is used to buy the house for the couple.

Three affluent women suffered from the messy financial situation of their male partners, including \textbf{debts and gambling}. In Post 9, one woman explains: ``\textit{He earned around 7,000 CNY per month. He used loans to support his luxurious lifestyle. I know he has loaned 200 thousand CNY, and he'll never be able to repay. But he never changes his lifestyle. He recruits prostitutes using thousands of CNY each time. He buys clothes without even looking at prices. He's very happy if a young lady calls him a boss or something. He just wants to establish himself as a successful figure outside.}'' Another woman explains: ``\textit{He's also addicted to gambling! He can lose hundreds of thousands of big bets in one single night, so many times his debts are repaid by my sister. He sometimes even beats my sister when he comes home after losing a gamble.}'' (Post 37).

Several types of \textbf{domestic violence}, or intimate partner violence (IPV), can be identified in the posts, including physical abuse (Posts 2, 37, 40, 77), verbal/emotional abuse (Posts 2, 55, 66, 77, 95), and control (Post 71). Post 77 is a typical example of verbal and emotional abuse. Her husband constantly belittled her: ``\textit{Whenever I told him funny jokes, he would say I was vulgar, and should read more books to improve myself and better educate our child. When he saw me watching TV series every day, he would accuse me of not spending my time on meaningful things, like making education plans for our child, and making plans for myself.}'' After her mother died, she was immersed in sadness. Again, her husband accused her of having a bad mentality and not being able to be a good role model for their child.

\textbf{Lack of emotional support/respect/romance} from their partners was also a common issue reported by women. In Post 29, the content contributor's husband showed no respect for her as well as other women when dining with his friends, but she was forced to show up on such occasions: ``\textit{I often sit there and listen to their rhetoric, mixed with swear words, sarcasm about female leaders, or complaint about the constraints that family brings to them. I feel that my gender has become very embarrassing and disrespectful. I'm like a decoration.}'' Some men seem to not be able to provide emotional value to their partners: ``\textit{My husband doesn't care about me at all, and I guess he doesn't even have the ability to do so. I feel that he doesn't know how to be around people, especially when it comes to intimate relationship. I don't demand anything from him now, and can only accept the reality. Divorce is not an option for me.}" (Post 46). Post 58 describes a similar situation. In Post 50, the husband was not aware how vulnerable pregnant women could be, and accused his pregnant wife of not doing as much housework. Another example for this category: ``\textit{Our family is like a patriarchal society. My father has a successful career, and of course his emotion is always given top priority at home. He's busy with work, socializes a lot, and rarely has time to have dinner at home and spend time with us. It's a great gift for us to have him at home.}'' (Post 86).

\textbf{Sex mismatch} seems to be accompanied with the husbands' troubled relation to sex. In Post 34, the content contributor's husband always refused to have sex with her. When she tried to communicate, her husband either refused to discuss it, or blamed everything on her, asking her to lose weight and get into good shape. While this woman saw the sex mismatch as an unsolvable issue, she bought herself sex toys to satisfy her desires and kinks, and acknowledged that she would not refuse to make love to other men if she met suitable ones. However, this is a rare case, and for other Chinese women, they ``have a harder time defending the legitimacy of their sexual desires than men.'' (Post 48).

\textbf{Concealing of illness} was described only in one case (Post 12): ``\textit{My brother is about to get married, and the wedding date has been set. He has always been a very good person in my mind. He makes money, does housework, and is tidy, but one of his kidneys is not very healthy. He was hospitalized before, but it has not been well till now. It probably means that it will affect his future sex life and fertility. However, he hasn't told his girlfriend about this. He, as well as our mom, is not going to disclose this to his fiance before getting married.}'' The author struggled between keeping the secret for her brother and telling the truth to her future sister-in-law.

\textbf{Work-family balance} is the decision and compromise many women have to make, especially when they work in different cities from their partners. The woman in Post 15 was confronted with such a situation: ``\textit{His family urge him to get married, and he wants me to move to his city. But I don't want to give up my career and go to a strange city to start over. Now we're in a semi-breakup state.}'' In Post 27, the woman's boyfriend agreed to marry her only after she quit her job in a company, passed the National Civil Servant Examination, and became a government official. According to the following post, the decision for women is made harder given the stereotypical measure of a woman's success in Chinese society: ``\textit{The pressure of being a woman in China and the oppression of a male-dominated society is too heavy. No matter how well women do their jobs, people still judge them by how successful their marriage is, or how early they get married. Many people are forced to get married, and to have a second child, and then a third child.}'' (Post 23).

Cheating behavior is heavily discussed in the posts. Nineteen posts were women's narration of their partners' or fathers' cheating behavior, which we categorize as \textbf{partner-cheating}. They often appear together (e.g., Posts 40, 41, 42, 44, 45), and given the temporal order of the posts, the content contributors may have been inspired by recent posts which stroke a response in them. Some directly indicated such a motivation: ``\textit{I just saw the post about parents' divorce. I'm shocked, because I've been treated exactly the same way.}'' (Post 66).

Eleven posts were women's narration of their own cheating behavior, which are categorized as \textbf{self-cheating}. They cheated for various reasons, from sex mismatch (Posts 42, 43) to retaliatory cheating (Post 59). The woman in post 77 had suffered from domestic violence and lack of emotional support and romance from her husband before she cheated. Her husband had also cheated before. Thus, she felt less of a moral burden or sense of guilt.

In \textbf{age anxiety/marry early}	cases, either the women want to marry early or their parents and partners urge them to, with the stereotypical, traditional assumption that young women are more popular on the marriage market. One woman is stuck between the choices of pursuing her degree abroad, or giving up her degree, going back to China, finding a job, and seeking a relationship and, subsequently, marriage. She thought at the age of 28, she ``\textit{basically had no future in the workplace and in the dating market.}'' (Post 36). Some parents urge their daughters to marry, even if their boyfriends have perceived bad traits: ``\textit{I think he's an extremely terrible individual. But in the eyes of my parents, he's a good candidate for marriage.}'' (Post 7). Another traditional mentality is that marriage is more important than career for women, which drives parents to urge their daughters to get married: ``\textit{Although my parents are happy with my achievements at work and think that career development is very important, they believe the most important thing for women is to get married and have children.}'' (Post 23). Some parents even decided when their daughters get married, regardless of their own will, as in the case of Post 8: ``\textit{In 2019, my parents introduced my boyfriend to me. We've been in a long-distance relationship and haven't spent much time together. I'm scared about marriage. But how can I break up with him? My parents have set a wedding date for us. I'm over 30, which also makes me anxious.}'' However, women who had suffered from their marriage still encouraged others to get married: ``\textit{My niece is around 30 years old, and my sister keeps urging her to get married. My niece and I, after seeing her two miserable marriages, have determined not to get married... It's like a dead end. A woman uses her own blood-and-tear experience to teach other women to step into the same river. I guess the bubble which has trapped many women for all their lives will finally burst, only after career-oriented women are recognized by society, and women no longer need to prove their value with their marriage.}'' (Post 37).

\textbf{Controlling parents} can be either the women's parents, or their partners' parents. The woman in Post 46 described how her mother-in-law kept bothering and criticizing her: ``\textit{If she doesn't like me, she'll just accuse me without hiding her feelings. All I can do is to endure it. I'm depressed, but I can do nothing. I can't afford to buy another house and live separately from her.}'' In the cases of Post 21 and 94, the parents interfere with and have the final say on every aspect of their children's life.

\subsection{Tolerance and self attribution}
In addition to the 20 categories of the presented issues, we also identified some mentalities which could worsen the women's situation.

\textbf{Tolerance} was sometimes found. This was observed if someone does not believe there are many potentially satisfactory male partners in China, so they have a relatively low standard when seeking a partner, as in the case of Post 36: ``\textit{I know men are all the same. A man who doesn't cheat and brag is already good enough. I just want someone who can talk with, play with, and who won't kill me when I sleep. All I want is to have a stable home and live an ordinary life.}'' When one woman was cheated on by her husband, her friends told her that men were basically the same: ``\textit{[After my husband cheated on me,] I consulted with my married friends, and it turned out that everyone's marriage was the same. They said I asked for too much, and all men would cheat. The only difference is if they're caught.}'' (Post 73).

\textbf{Self attribution} is when the women ascribe the issues happening in their relationship only to themselves. For example, in Post 35, the woman and her boyfriend struggled with the distribution of housework as he suggested that it was only a woman's responsibility. Consequently, she attributes their breakup to herself: ``\textit{Most problems in our relationship are because of me. I wasn't able to solve them well at that time, and I also missed the opportunity to learn how to solve problems and communicate, which is a huge regret for me.}''

\section{Discussion}
For some cultures, women are willing to share their negative experiences non-anonymously, as was seen in the \#MeToo movements in western countries \cite{introduction:1, introduction:2}. For China, however, this can be different. Influenced by Confucian mentality, cultural norms and social sanctions regarding sex and sexual harassment are issues people are ashamed to talk about \cite{metoo:2}. Here, cultural aspects like face-saving and associated shame can make it harder for women to express such problems \cite{weibo:1}. 

Research about anonymity on social media suggests that people seem to be freer in their expressions, potentially also leading to harassment and bullying \cite{anonymous:8}. By being able to express oneself freely without fear of judgement, people could be more willing to share their own experiences \cite{anonymous:5}. As such, sensitive topics such as intimate partner violence, workplace harassment or cheating behavior can be openly talked about \cite{weibo:1, introduction:1}. 

Recently, feminist influencers on Sina Weibo started anonymously posting incidences of affected women on their social media channels. With this study, a content analysis was conducted to investigate what types of gender issues are shared anonymously by women through a Chinese influencer, called Qiaomai. To this end, 774 posts were collected from January to March 2022 of which only 100 were analyzed due to time constraints. There is a lack of literature about anonymous expression of Chinese women on social media, therefore, the conventional approach for content analysis was applied. 

All in all, 20 categories of gender issues were defined through observation of the posts, consisting of such topics as ambiguous relationship, harassment, and domestic violence. Each post was coded by two independently working coders who discussed their results afterwards, resulting in a Krippendorff's alpha of at least 0.663 for all variables \cite{method:2}. The most frequently observed categories deal with cheating-partners, followed by controlling parents, and age anxiety/marrying early. Positive experiences are also shared, e.g., after a divorce from a toxic marriage. Compared to existing literature, our analysis uncovered more gender/social issues expressed by women on social media, in addition to IPV \cite{weibo:1}. 

The observed gender issues anonymously expressed by women are contextualized in a culture which has long been known to be conservative, where women's value is attached to their family instead of their own achievements, and where women's rights including reproductive freedom are not well protected \cite{china:1, china:2}. What's worse, when confronted with these gender issues, they may find difficulty in seeking supportive resources, due to the limited social support and a lack of psychological professionals. According to Fang et al., China has roughly 5,000 clinical psychologists, only 1/4 of the number in the United States, for a population four times as big. Also, only a small number of economically well-developed areas in China have social workers serving mental health patients \cite{mental}. By sharing their experiences anonymously with influencers, empathy, online interpersonal support and practical advice could be made accessible to the affected women. The comments also allow like-minded women to connect with each other. As we notice, posts of the same topic often appear together, indicating that when a gender issue is publicly expressed, it may arouse similar discussions.  

This study's observations echo with previous research which finds that anonymity can encourage expressiveness and interaction among users, and make the disclosure of deep confessions and secrets possible \cite{anonymous:7}. In a culture where face-saving is seen important \cite{weibo:1}, the women in this study nevertheless express their personal experiences, which were mostly negative.

While many communities for women have focused on daily life \cite{community:2} and entertainment \cite{community:1}, this study revealed the role of an online community to help women express their negative experiences.

\subsection{Limitations and future work}
To this end, an exploratory understanding of common gender issues anonymously shared by female Weibo users was provided. A follow-up study can further analyze the dynamics of the discussion around such issues, e.g., if the affected women receive proper help in comments, if moral judgments exist especially when it comes to traditionally taboo subjects such as women's self-disclosure of their cheating behavior. 

Given the specific site for observation (Sina Weibo), the findings may not be generalized to other cultural contexts. Future research could consider a cross-cultural analysis of gender issues anonymously expressed on social media, e.g., comparing between Weibo and Reddit. 

Last but not least, the authors find that the anonymous experiences are mostly shared by well-educated women, with good expression skills. Women with a lower educational level or from rural areas may not have the ability or Internet access to express themselves online. As indicated in \cite{metoo:3}, rural and working-class women are largely marginalized and underrepresented in China's present feminist movements. Future research could take marginalized women into consideration.

\section{Conclusion}
With a qualitative content analysis, we gained an in-depth understanding of how women anonymously expressed gender issues, from domestic violence to cheating, on a Chinese social media platform. These issues are sometimes associated with mentalities that could be seen as more traditional, such as misinterpretation of women's value (marry early), and gender stereotypes (lack of reproductive freedom). We sometimes find women to attribute relationship issues to themselves, as well as to tolerate their partners' misbehavior. The initial taxonomy can potentially help policy makers in providing opportunities for women to share their experiences, and encourage more study on this subject.


\bibliographystyle{unsrt}  
\bibliography{references}  

\end{document}